\documentclass{raa}            
\pdfoutput=1
\usepackage{graphicx,times}  
\usepackage{natbib}
\usepackage{amssymb,amsmath}
\usepackage{amsmath}
\usepackage{multirow}
\usepackage{tabularx}
\usepackage{booktabs}
\usepackage[flushleft]{threeparttable}
\usepackage{lipsum}
\usepackage{enumitem}
\usepackage{hyperref}
\usepackage{siunitx}
\usepackage{rotating}
\usepackage{adjustbox} 
\usepackage{longtable}
\usepackage{xltabular}
\usepackage{xcolor}
\usepackage{textcomp}
\bibpunct{(}{)}{;}{a}{}{,}

\usepackage{tikz,xcolor,hyperref}

\definecolor{lime}{HTML}{A6CE39}
\DeclareRobustCommand{\orcidicon}{%
	\begin{tikzpicture}
		\draw[lime, fill=lime] (0,0) 
		circle [radius=0.16] 
		node[white] {{\fontfamily{qag}\selectfont \tiny ID}};
		\draw[white, fill=white] (-0.0625,0.095) 
		circle [radius=0.007];
	\end{tikzpicture}
	\hspace{-2mm}
}

\foreach \x in {A, ..., Z}{%
	\expandafter\xdef\csname orcid\x\endcsname{\noexpand\href{https://orcid.org/\csname orcidauthor\x\endcsname}{\noexpand\orcidicon}}
}

\begin{document}
	
	\title{The effect of external magnetic field on electron scale Kelvin-Helmholtz instability.	}
	
	\volnopage{Vol.0 (20xx) No.0, 000--000}      
	\setcounter{page}{1}          
	
	\author{D. Tsiklauri
		\inst{1}\orcidA{}}

	\institute{Joule Physics Laboratory, School of Science, Engineering and Environment, 
University of Salford, Manchester, M5 4WT, United Kingdom, {\it D.Tsiklauri@salford.ac.uk}\\
	\vs\no
		{\small Received 2024 July 18; revised 2024 August 13;
		accepted 2024 August 24}}
	
	\abstract{We use particle-in-cell, fully electromagnetic, plasma kinetic 
simulation to study
the effect of external magnetic field 
on electron scale Kelvin-Helmholtz instability (ESKHI). 
The results are applicable to collisionless plasmas when e.g.
solar wind interacts with planetary magnetospheres or 
magnetic field is generated in AGN jets.
We find that as in the case of  magnetohydrodynamic KHI,
in the kinetic regime, presence of external magnetic 
field reduces growth rate of the instability.
In MHD case there is known threshold 
magnetic field for KHI stabilization, while for ESKHI
this is to be analytically determined. 
Without a kinetic analytical expression, 
we use several numerical simulation runs to establish an empirical
dependence of ESKHI growth rate, $\Gamma(B_0)\omega_{\rm pe}$, 
on the strength of applied external magnetic field. 
We find the best fit is hyperbolic, 
$\Gamma(B_0)\omega_{\rm pe}=\Gamma_0\omega_{\rm pe}/(A+B\bar B_0)$, where $\Gamma_0$ is the 
ESKHI growth rate without external magnetic field
and $\bar B_0=B_0/B_{\rm MHD}$ is the ratio of
external and  two-fluid MHD stability threshold magnetic field, derived here.
An analytical theory to back up this growth rate dependence on
external magnetic field is needed.
The results suggest that in astrophysical settings where
strong magnetic field pre-exists, the generation
of an additional magnetic field by the ESKHI is
suppressed, which implies that the Nature provides a 
"safety valve" -- natural protection 
not to "over-generate" magnetic field by ESKHI mechanism.
Remarkably, we find that our two-fluid MHD threshold magnetic field 
is the same (up to a factor $\sqrt{\gamma_0}$)
as the DC saturation magnetic field, previously predicted by fully kinetic theory. 
\keywords{instabilities -- magnetic fields -- plasmas -- 
Sun: heliosphere -- ISM: magnetic fields}
	}
	\authorrunning{Tsiklauri }            
	\titlerunning{Electron scale Kelvin-Helmholtz instability}  
	\maketitle
\section{Introduction}

Electron scale Kelvin-Helmholtz instability is a relatively
new offspring \citep{gruzinov2008,Alves_2012,Grismayer_2013,Alves_2014,Alves_2015,Miller_Rogers_2016,Yao_2020,M_2020} of the classical version the instability
 that has been known since 1868 \citep{hh1868,lk1871}.
The instability occurs due to a velocity difference (i.e. shear) 
at an interface. The interface can be between either
(i)  different parts of a single fluid or 
(ii) two different media, as long
as there is said velocity difference.
As this is a mature field of research, vast amounts of literature
exist on the subject. The instability can occur in the media that is
(i) magnetized or unmagnetized i.e. with or without
external magnetic field.
(ii) Also, the description and the properties of the instability
significantly differ in kinetic or fluid-like regimes.
We mostly focus our consideration on kinetic description of
the instability.

In general, there are two contexts for this study: 
collisionless plasmas found in situations were
(i) solar wind interacts with planetary magnetospheres 
\citep{delamere_2011, Foullon_2011, Johnson_2014, Delamere_2021} or 
(ii) when magnetic field is generated in astrophysical scenarios, such as
active galactic nucleus and gamma-ray bursts \citep{gruzinov2008,Alves_2012,Grismayer_2013,Alves_2014,Alves_2015}.

Accepted models of Gamma Ray Bursts rely
on the presence of background magnetic field.
It appears that magnetic field energy and
kinetic energy of the accelerated particles are in equipartition. 
This implies that aforesaid magnetic
field needs to be somehow generated \citep{gruzinov2008}. 
\citet{Alves_2012} presented
the first self-consistent three-dimensional 
particle-in-cell (PIC) simulations of ESKHI.
The main findings of this work include
establishing the saturation levels of
maximum equipartition values of 
$E_B /E_p \approx {\rm few} 
\times  10^{-3}$, where $E_B=\int B^2/(2\mu_0) dV$ and 
$E_p=\int \rho v^2/2 dV$ are the volume integrated 
magnetic and particle kinetic energy densities, respectively. 
\citet{Alves_2012} found what factors prescribe
the level of saturation of the magnetic field
generated by the ESKHI, which typically
occurs on electron scales i.e. circa
100 electron plasma periods.
Set up of \citet{Alves_2012}, which considers
a regime of equal speed electron counter-flow layers of 
equal number density across the interface,
is relevant to GRB shocks \citep{Piran_2005}, 
where density shells have similar number densities and
the relativistic factor is in the range
$1 \leq \gamma_0 \leq 10$.
Particle-in-cell numerical simulations presented
by \citet{Alves_2014} established the generation 
of a sizable, $E_B /E_p \approx {\rm few} \times  10^{-3}$,
 and large-scale DC magnetic field component, 
not predicted by a linear, fluid-like
description and only appears in the kinetic regime. 
\citet{Alves_2014} showed that the generated magnetic field
appears due to
thermal expansion of electrons of one flow into
the other across the shear interface. 
At the same time, in \citet{Alves_2014} ions stay motionless due to the large
mass. This electron expansion was found to form 
current sheets, which generates the magnetic field.
\citet{Alves_2014} extended previous work by \citet{gruzinov2008},
by considering different number densities across the
shear flow interface and derived a new dispersion
relation.
\citet{Alves_2014} also considered smooth 
shear flow profiles such as
$v_0 (x)/ c = 0.2 \tanh (x /L )$
and found that smoother shears produce smaller
ESKHI growth rates.
\citet{Alves_2014} this way provided
a generalization of MHD result by \citet{Miura_1982},
who in turn found that, in the compressible case
the growth rate is a function of the magnetic Mach number and
the modes with $k L < 2$ are unstable.
\citet{Miura_1982} also found that the
most unstable modes have their 
wavelength comparable to the width of the shear layer
 $2 k L \simeq 1$.
A very important distinction between kinetic and MHD regimes
is underscored by \citet{Alves_2014}, on page 9,  which we quote
without an alteration: "{\it shear flow instabilities in initially unmagnetized
conditions with fast drift velocities (relative to the temperature) can only develop on the
electron-scale}".
This underscores the importance of kinetic effects which is
one of the main motivations for this study in the KHI context.
\citet{Alves_2015} studied a new
type of kinetic instability, so-called
mushroom instability (MI), named so because of
the mushroom-shaped features found in the
electron number density.
The difference between ESKHI and MI is that
for ESKHI $\vec v_0(x) \parallel Oy$, while
for ESKHI $\vec v_0(x) \parallel Oz$.
\citet{Alves_2015} studied how the 
growth rates of ESKHI and MI scale with 
$\beta_0=v_0/c$ and $\gamma_0$ and found that 
the ESKHI has higher growth rates than the
MI for sub-relativistic settings. 
However, the MI growth rate
decays with $\gamma_0^{-1/2}$, 
slower than the ESKHI, which decays with
$\gamma_0^{-3/2}$ 
(see Fig. 1 from \citet{Alves_2015}). Thus they concluded that
ESKHI dominates for $\gamma_0 \approx 1$ for sub-relativistic flows,
while MI dominates for $\gamma_0 \gg 1$.
Because of this reason, i.e. that we would like
our results to be applicable to both 
(i) collisionless plasmas when
solar wind interacts with planetary magnetospheres or 
(ii) magnetic field is generated in places, such as
active galactic nucleus and gamma-ray bursts with relatively moderate
$\gamma_0$'s, this paper focuses mostly ESKHI with $\beta_0=v_0/c =0.2$ 
($\gamma_0=1.02$) -- this is the value also 
considered by \citet{Alves_2014}.

We mention in passing that, while the above discussion was
for the electron-proton plasmas, a body of work exists
on electron-scale kinetic, relativistic shear instabilities,
where similarly, the  magnetic field generation is seen but
in electron-positron plasmas
\citep{Liang_2013}. A comparison of the electron-positron results to electron-proton
plasmas \citep{Liang_2013b,Nishikawa_2014} or dependence 
of the growth rate on the ion-to-electron mass ratio
has been also studied \citep{Nishikawa_2013}.

\citet{Miller_Rogers_2016} extended analysis of
\citep{gruzinov2008,Alves_2012,Alves_2014,Alves_2015}
by considering a warm plasmas and found
that the growth rate is significantly, up to a factor of 3,
larger for the case of large temperatures.
This analytical calculation conclusion by \citet{Miller_Rogers_2016}
is supported by the  multi-dimensional particle-in-cell simulations 
of \citet{Grismayer_2013}.
\citet{Yao_2020}  also analyzed the role of electron thermal motion 
effects on the generation of the magnetic field.
\citet{Yao_2020} found an increase in the 
growth rate with the increasing plasma temperature. 
\citet{M_2020} studied the instability growth rate of the excited electromagnetic modes for the relativistic and non-relativistic
cases of solar wind, interacting with interstellar plasma medium
with the emphasis of the effect of the viscosity of plasma.

The above introductory comments were all in the kinetic regime.
In the  fluid-like description of KHI, 
the first paper which considered the effect of
external magnetic field on KHI was \citet{Michael_1955}.
It should be noted, it  was \citet{Michael_1955}
who first derived the dispersion relation for
the simple case of an incompressible plasma with a
discontinuous flow shear with the
perturbations to the interface between
of two conducting media, with velocities 
$U_0$ and $U_1$ and constant magnetic fields $B_0$
and $B_1$ that are parallel to the interface.
Many published works wrongly attribute this result to
\citet{Chandrasekhar_1961}, which is a later work.
\citet{Blandford_1976} studied
linearized Kelvin-Helmholtz instability with a
calculation that generalized previous 
treatments to include relativistic 
relative motion and relativistic 
internal sound speeds. The study was performed
in the context of beam models of extra-galactic 
radio sources. 

The motivation for the present work is two-fold:
(i) To extend MHD analysis of \citet{Michael_1955} to the kinetic
regime of ESKHI; and
(ii) To extend kinetic analysis of \citet{Alves_2014} 
by adding the effect of external magnetic field to ESKHI.
 
The paper is organized as following: Section 1 
provides an introduction to the subject of ESKHI.
Section 2 discusses prior analytical and numerical
findings about ESKHI.
Section 3 provides the details of our model.
Section 4 reveals  the main results of this study.
Section 5 lists the main conclusions of this work. 

\section{Prior analytical and numerical
findings about ESKHI}

 An analytic 
calculation by \citet{gruzinov2008} 
provides the growth rate of ESKHI in 2D.
A limited, relevant number of components of 
background relativistic shear flows and  number densities
\begin{equation}
\vec v_0=(0,v_0(x),0), \;\;\; n_0=n_0(x)
\label{eq1},
\end{equation} 
as well as electromagnetic
field perturbations 
\begin{equation}
\vec E_1=(E_{1x}(x),E_{1y}(x),0), \;\;\;  \vec B_1=(0,0,B_{1z}(x)),
\label{eq2}
\end{equation} 
with the perturbation wave-vectors having only y-component and
harmonic time dependence as $f_1=\tilde f_1(x)e^{i (k_y y-\omega t)}$
were considered.
\citet{gruzinov2008}
established that for equal speed electron counter-flow layers
of the same number density across the interface i.e. for
$v_0(x)=v_0 {\rm sign}(x)$ and $n_0(x)=const$,
the growth rate is
\begin{equation}
\Gamma_0^2=\frac{\omega_{\rm pe}^2}{2}\left(
\sqrt{1+8\frac{k^2v_0^2}{\omega_{\rm pe}^2}}
-1-2\frac{k^2v_0^2}{\omega_{\rm pe}^2}\right).
\label{eq3}
\end{equation}
Eq. \ref{eq3} then implies that the
condition for the ESKHI is $|k v_0| < \omega_{\rm pe}$.
Figure 3 from \citet{Alves_2014}
gives a graphical representation 
of the growth rate of ESKHI versus wave number.
It appears like an up-side-down parabola, with
skewed to the right  maximal growth rate of
\begin{equation}
\Gamma_0 =\frac{\omega_{\rm pe}}{2\sqrt{2}} \approx 0.35\omega_{\rm pe}
\label{eq4}
\end{equation}
at the most unstable wave-number
\begin{equation}
k v_0=\frac{\sqrt{3} \omega_{\rm pe}}{2\sqrt{2} } \approx 0.61 \omega_{\rm pe}.
\label{eq5}
\end{equation}
Note that in the above equations $k$ and $\omega_{\rm pe}$ include
 the relativistic factor 
 $\gamma_0=1/\sqrt{1-v_0^2/c^2}$ dependence.
While, \citet{Grismayer_2013} give a useful, explicit dependence on 
the relativistic factor $\gamma_0$, and draw a distinction
between $k_\perp$ and $k_\parallel$ of the form
$f_1=\tilde f_1(x)e^{-k_\perp |x|}e^{i(k_\parallel y - \omega t)}$:
\begin{equation}
\Gamma_0^2=\frac{\omega_{\rm pe}^2}{2 \gamma_0^3}\left(
\sqrt{1+8\frac{k_\parallel^2v_0^2\gamma_0^3}{\omega_{\rm pe}^2}}
-1-2\frac{k_\parallel^2v_0^2\gamma_0^3}{\omega_{\rm pe}^2}\right),
\label{eq6}
\end{equation}
\begin{equation}
\Gamma_{0,\rm max} =\frac{1}{2\sqrt{2}}\gamma_0^{-3/2}\omega_{\rm pe} \approx 0.35 \gamma_0^{-3/2} \omega_{\rm pe},
\label{eq7}
\end{equation}
\begin{equation}
k_{\parallel, \rm max} v_0=\frac{\sqrt{3} \gamma_0^{-3/2}\omega_{\rm pe}}{2\sqrt{2} } \approx 0.61 \gamma_0^{-3/2} \omega_{\rm pe},
\label{eq8}
\end{equation}
with $\omega_{\rm pe}=\sqrt{n_e e^2/(m_e \varepsilon_0)}$ 
being electron plasma frequency for $\gamma_0=1.0$, strictly.
  
\citet{Alves_2012} performed self-consistent three-dimensional 
particle-in-cell simulations 
to study ESKHI.
They found that the saturation levels of
 maximum equipartition values are 
 $E_B /E_p \approx 2 \times  10^{-3}$
for the sub-relativistic scenario, 
and $E_B /E_p \approx 7 \times  10^{-3}$ for
the relativistic scenario. In their terminology
sub-relativistic means $\gamma_0 = 1.02$ 
(i.e. $v_0/c=\sqrt{1-1/\gamma_0^2}=0.1971\approx0.2$) 
and sub-relativistic means $\gamma_0 = 3$ 
(i.e. $v_0/c=\sqrt{1-1/\gamma_0^2}=0.9428\approx0.9$).
Also, \citet{Alves_2012} and \citet{Grismayer_2013} established what prescribes
the level of saturation of the magnetic field
generated by the ESKHI, which typically
occurs on "electron scales" circa 
$\approx 100 /\omega_{\rm pe}$
with the saturation magnetic field
given by
\begin{equation}
B_{\rm DC}\simeq  \frac{m_e \omega_{\rm pe}}{e} \beta_0 \sqrt{\gamma_0}.
\label{eq9}
\end{equation}
Note that in Eq. \ref{eq9} $B_{\rm DC}$
is in SI units while \citet{Grismayer_2013} uses CGS, hence
conversion of magnetic field and charge yields a factor of
$\sqrt{\mu_0/ 4 \pi} \times \sqrt{4 \pi \varepsilon_0}=1/c$.

\section{Description of the model}
\subsection{Theoretical considerations}

Because our motivation for the present work is, on one hand, to extend MHD analysis of \citet{Michael_1955} to the kinetic
regime of ESKHI and, on the other hand,
to extend kinetic analysis of \citet{Alves_2014} 
by adding the effect of external magnetic field to ESKHI,
we need to somehow fix the relevant magnetic field scale.
It is a common knowledge that usually in
many space and astrophysical plasma situations
"MHD works where it should not", so in the absence
of an analytical theory of ESKHI with an external
magnetic field, we fix the relevant magnetic field 
scale as two-fluid
 MHD stability threshold magnetic field, based on the calculation
given in Appendix A.
In particular \citet{Michael_1955}'s dispersion relation for
an incompressible plasma with a
discontinuous flow shear with the
perturbations to the interface between
of two conducting media, with velocities 
$U_0$ and $U_1$ and constant magnetic fields $B_0$
and $B_1$ that are parallel to the interface, reads
as
\begin{equation}
 \frac{\omega}{k}=-\frac{U_1+U_2}{2}\pm \sqrt{ \frac{(B_1^2+B_2^2)}{2 \mu_0 \rho}-\frac{\Delta U^2}{4}},
\label{eq10}
\end{equation}
where $\Delta U=U_1-U_2$ is the flow velocity difference
across the interface.
Eq. \ref{eq10} suggests then the
existence of stability threshold magnetic fields that satisfy
\begin{equation}
  \frac{(B_1^2+B_2^2)}{2 \mu_0 \rho}=\frac{\Delta U^2}{4}.
\label{eq11}
\end{equation}
Special cases are:
(i) a case without external magnetic field $B_1=B_2=0$ that recovers
KH result that the current sheet is always unstable as long as there 
is a velocity difference; and
(ii) a case of $U_1=-U_2=v_0$ and $B_1=B_2=B_0$ considered in this paper,
then the stability threshold magnetic field is
\begin{equation}
  \frac{B_0^2}{2 \mu_0 }=\rho \frac{v_0^2}{2},
\label{eq12}
\end{equation}
which physically means that threshold magnetic field
is achieved when magnetic field energy density is equal
to counter-flow kinetic energy density.

In the kinetic description, the electron dynamics is crucial,
while massive ions essentially provide neutralizing background.
For example \citet{Alves_2014} calculation pertains to
moving electrons only.
We therefore make an important change to Eq. \ref{eq12},
namely, $\rho \to \rho_e= n_e m_e$. Which means
to switch  to two-fluid MHD with $\vec v\to\vec v_e$, i.e.
opposite to a single fluid MHD in which electrons and ions are
"glued" to each other and act as a single fluid.
Thus, for the purposes on this paper, based on
Eq. \ref{eq12}, we set 
\begin{equation}
B_{\rm MHD}=\sqrt{\mu_0 m_e n_e} v_0.
\label{eq13}
\end{equation}

Using the definition of electron
plasma frequency $\omega_{\rm pe}=\sqrt{n_e e^2/(m_e \varepsilon_0)}$
and the expression $c=1/\sqrt{\mu_0 \varepsilon_0}$,
the Eq. \ref{eq13}
can be rewritten in the notations of
Eq. \ref{eq9} as:
\begin{equation}
B_{\rm MHD}= \frac{m_e  \omega_{\rm pe}}{e} \left(\frac{v_0}{c}\right)=
 \frac{m_e  \omega_{\rm pe}}{e} \beta_0.
\label{eq14}
\end{equation}
It is remarkable that \citet{Grismayer_2013}'s Eq. \ref{eq9} and 
our \ref{eq14}
are the same, up to a factor of $\sqrt{\gamma_0}$.
In other words,  two-fluid MHD
threshold magnetic field, derived here -- see Appendix A for derivation
of the extended version of Eq. \ref{eq10} --
which suppresses
the KH instability, is the same (up to a factor of $\sqrt{\gamma_0}$
which is 1.02 for the present paper)
as the DC saturation magnetic field
predicted by the kinetic theory and simulations of \citet{Grismayer_2013}.

We specifically refrain from using term electron MHD or EMHD
which has a more specific meaning. 
Instead we refer to "two fluid MHD",
or it would be more precise, but {cumbersome} to
use term "two-fluid MHD with stationary ions".
The following discussion explains as to why:
According to \citet{Lyutikov_2013},
who in turn, basis his discussion on \citet{Gordeev_1994},
within the framework of EMHD, entire 
electric current is carried by electron fluid:
$\vec J = -e n_e \vec v_e$, ions only provide a neutralizing background
and do not move, i.e. do not contribute to pressure or
mass (inertia).
For the case of
infinite conductivity, 
the magnetic field is frozen into into
electron fluid and thus an electric field 
satisfies the condition,
\begin{equation}
\vec E + \vec v_e \times \vec B = 0.
\label{eq15}
\end{equation}
Hence the magnetic field induction equation is of the form
\begin{equation}
\frac{\partial \vec B}{\partial t}=
\nabla \times (\vec v_e \times \vec B).
\label{eq16}
\end{equation}
In EMHD approximation, the next crucial step is
replacing $\vec v_e$ in the Eq. \ref{eq16} using 
the current expression $\vec v_e = - \vec J /(e n_e)$,
where $\vec J = \nabla \times \vec B / \mu_0 $, i.e.
$\vec v_e = - \nabla \times \vec B /(\mu_0  e n_e)$:
\begin{equation}
\frac{\partial \vec B}{\partial t}=-\frac{1}{\mu_0  e}
\nabla \times \left[  \frac{(\nabla \times \vec B)}{n_e} \times \vec B \right].
\label{eq17}
\end{equation}
\citet{Cho_2009} provide a rigorous 
physical interpretation for the EMHD approximation, by
considering MHD, Hall MHD and EMHD at appropriate
spatial scales. They use ion inertial length,
$d_i = c / \omega_{\rm pi}$, for their ordering
of terms (see discussion around their Equations
8--13). We instead use electron inertial length,
$d_e = c / \omega_{\rm pe}$.
In the Hall MHD approximation one has 
\begin{equation}
\frac{\partial \vec B}{\partial t}=\nabla \times (\vec v \times \vec B)-\frac{1}{\mu_0  e}
\nabla \times \left[  \frac{(\nabla \times \vec B)}{n_e} \times \vec B \right].
\label{eq18}
\end{equation}
Note that in Eq. \ref{eq18}
the first term contains the bulk plasma velocity $\vec v$, not the electron one
$\vec v_e$.
In Eqs. \ref{eq17} and \ref{eq18}
the factor $1/(\mu_0 e n_e)$ can be written as
$1/(\mu_0 e n_e)= d_e^2 e /m_e$.
Now expressing $e /m_e$ using Eq. \ref{eq14}
we get $1/(\mu_0 e n_e)= d_e^2 \omega_{\rm pe} v_0 /(B_0 c)$.
Hence with the substitution in Eq. \ref{eq18} of
$\vec v \simeq v_0$, $\nabla \simeq 1/L_0$,
and $\nabla  t \simeq 1/T_0$, the different term ordering is:
\begin{equation}
\frac{B_0}{T_0} \simeq \frac{B_0 v_0}{L_0}- \left(\frac{d_e}{L_0}\right)
\frac{B_0 v_0}{L_0}.
\label{eq19}
\end{equation}
Thus, from Eq. \ref{eq19} it follows that
for spatial scales much larger than electron inertial
length $L_0 \gg d_e$ usual single fluid MHD applies
while on scales $d_e \ll L_0$ EMHD applies.
But ESKHI is not at that scale, it is at a scale of
$L_0 \simeq d_e$. That is why we refrain from the use of 
term EMHD.

We also mention a useful notation of the magnetic field
induction in EMHD approximation provided by
\citet{Zhao_2010}:
\begin{equation}
\frac{\partial(1-d_e^2\nabla^2 )\vec B}{\partial t}=
\nabla \times (\vec v_e \times (1-d_e^2\nabla^2 )\vec B).
\label{eq20}
\end{equation}
Note that Eq. \ref{eq20} reduces to Eq. \ref{eq16}
with $d_e^2\nabla^2=d_e^2/L_0^2 \ll 1$.

With this distinction clearly stated, we refrain from 
using the induction equation of the form
Eq. \ref{eq17}, we stick with Eq. \ref{eq16}
rewritten as in \citet{Michael_1955}.
In the Appendix A we provide a calculation similar
to \citet{Michael_1955}, but with different densities on the
either side of shear interface and with the
substitution $\vec v\to\vec v_e$ and $\rho\to\rho_{e}$, since only
electrons move and stationary ions cannot contribute to
the mass (inertia).
We now state the main, starting equations of two-fluid MHD with stationary ions:
\begin{equation}
\frac{\partial \vec B}{\partial t}+
 (\vec v_e \cdot \nabla) \vec B =
(\vec B \cdot \nabla) \vec v_e ,
\label{eq21}
\end{equation}
\begin{equation}
\frac{\partial \vec v_e}{\partial t}+
(\vec v_e \cdot \nabla) \vec v_e= -\frac{\nabla p_e}{\rho_e}
+\frac{(\nabla \times \vec B) \times \vec B}{\mu_0 \rho_e},
\label{eq22}
\end{equation}
\begin{equation}
\nabla \cdot \vec v_e=0, \;\;\; \nabla \cdot \vec B=0.
\label{eq23}
\end{equation}
In the Appendix A we derive two-fluid MHD dispersion relation
for the different electron densities, $\rho_{e1}$ and $\rho_{e2}$, on
the either side of the interface, where we obtain:
\begin{equation}
 \frac{\omega}{k}=-\frac{U_{e1}\rho_{e1}+U_{e2}\rho_{e2}}{(\rho_{e1}+\rho_{e2})}
 \pm \sqrt{ \frac{(B_1^2+B_2^2)}{ \mu_0 (\rho_{e1}+\rho_{e2})}-\frac{\rho_{e1}\rho_{e2}\Delta U_e^2}{(\rho_{e1}+\rho_{e2})^2}},
\label{eq24}
\end{equation}
where $\Delta U_e=U_{e1}-U_{e2}$.

For the same electron density $\rho_e$ on either side of the interface we obtain:
\begin{equation}
 \frac{\omega}{k}=-\frac{U_{e1}+U_{e2}}{2}\pm \sqrt{ \frac{(B_1^2+B_2^2)}{2 \mu_0 \rho_e}-\frac{\Delta U_e^2}{4}},
\label{eq25}
\end{equation}
 and further still with 
$U_{e1}=-U_{e2}=v_0$, i.e.
$\Delta U_e=2 v_0$ and $B_1=B_2=B_0$ considered in this paper,
we obtain:
\begin{equation}
 \frac{\omega}{k}=0\pm \sqrt{ \frac{B_0^2}{\mu_0 \rho_e}-\frac{\Delta U_e^2}{4}}.
\label{eq26}
\end{equation}
Defining the {\it electron}-scale Alfven speed as
$V_{Ae}={B_0}/\sqrt{\mu_0 \rho_e}$,
then according to Eq. \ref{eq26} the
stability threshold is
\begin{equation}
 \Delta U_e > 2 V_{Ae} \;\;\; {\rm or} \;\;\; v_0> V_{Ae}.
\label{eq27}
\end{equation}
We also note that for the above conditions, the real part of
frequency is zero $\omega_r=0$.

A word of caution is that 
we are fully aware of the fact that Eqs. \ref{eq21}-\ref{eq23}
ignore the the displacement current which means that
relativistic effects are ignored because in fluid theory
ratio of electric field and magnetic forces are of the order
$(|\rho_e \vec E|/|\vec j \times \vec B|) \simeq v_0^2/c^2$. The numerical runs presented in this paper
are for $v_0^2/c^2=0.2^2=0.04 \ll 1$ so the above equations are
valid. We are also aware that a proper calculation
should follow similar approach as \citet{Alves_2014,Alves_2015}
where electric field (the displacement current) 
effect is explicitly present.
Hence, we remark that an
analytical theory to back up the growth rate dependence on
external magnetic field is outstanding and still needed.
We stress that, we only use the calculations based on
Eqs. \ref{eq21}-\ref{eq23}, and their outcomes 
just to set the scale of 
two-fluid (non-relativistic) MHD stability threshold magnetic field, given by Eqs. \ref{eq13} and \ref{eq14}.

\begin{figure*}
\captionsetup{justification=raggedright,
singlelinecheck=false}
\begin{center}
  \makebox[\textwidth]{\includegraphics[width=0.99\textwidth]{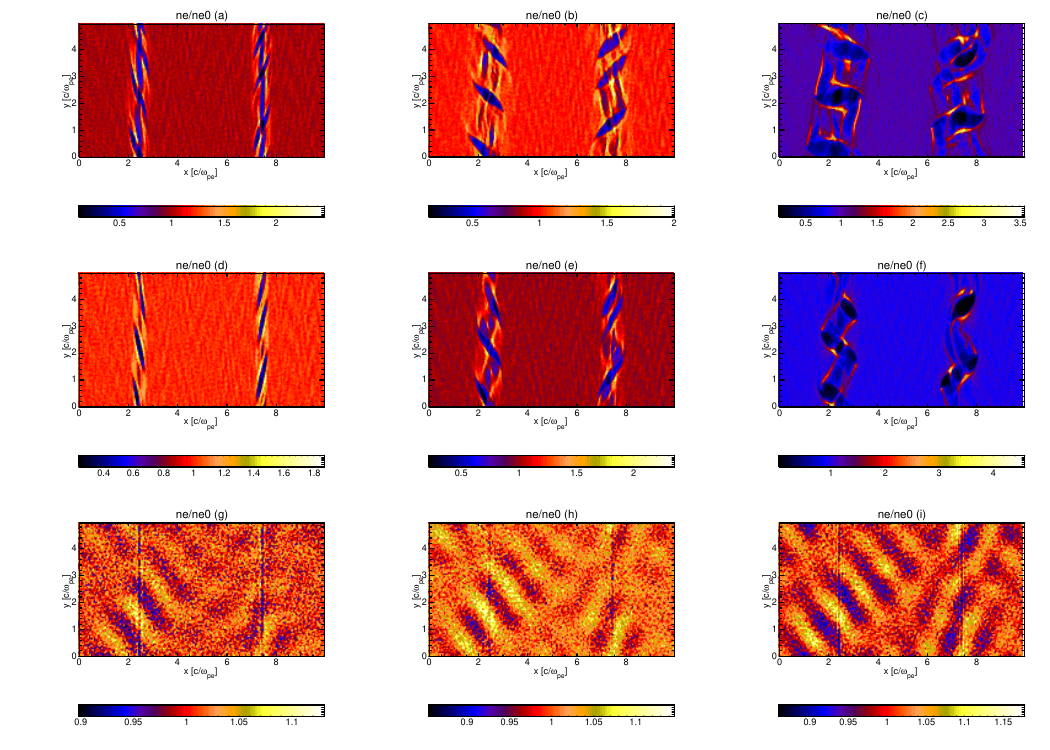}}
\end{center}
\caption{Snapshots of electron
number density 
$n_{\rm e}$.
The row of panels (a), (b) and (c) show $n_{\rm e}/n_0$  for
Run 0 at times $t=20, \; 25, \; 30/\omega_{\rm pe}$.
The row of panels (d), (e) and (f) show $n_{\rm e}/n_0$ for
Run 2 at times $t=20, \; 25, \; 30/\omega_{\rm pe}$.
The row of panels (g), (h) and (i) show $n_{\rm e}/n_0$ for
Run 6 at twice the times $t=40, \; 50, \; 60/\omega_{\rm pe}$. See Table \ref{t1} for details.}
    \label{fig1}
\end{figure*}

\subsection{Numerical simulation considerations}
In this work we use 2.5D version of PIC code EPOCH.
This a 1.5, 2.5 and 3D, fully electromagnetic, PIC code
\citep{Arber:2015hc}.
In order to be able to use periodic boundary conditions
throughout, which we remark, are the most precise boundary
condition for the numerical implementation,
we use a "sandwich" with three layers of plasma with the following
properties:
two down-flow layers
satisfying $x/L_x <0.25$ or $x/L_x>0.75$ with velocity $v_0/c=-0.2$
and one up-flow layer with $0.25< x/L_x<0.75$ with velocity $v_0/c=0.2$
in between and $v_0$ being parallel to $Oy$ axis.
The system size is $(L_x,L_y)=(10 c/ \omega_{\rm pe}, 5 c/ \omega_{\rm pe})$
which is resolved with $200\times100$ grid.
This means the each grid cell has a size of
1/20th of $c/ \omega_{\rm pe}$, i.e. 
$\Delta_x=\Delta_y=0.05 c/ \omega_{\rm pe}$.
Maximal temperature is set as $T_{\rm e,i, max}=m_e(0.005c)^2/k_B=
148239.77$ K. This ensures a cold plasma approximation
because $v_{\rm th,e}=\sqrt{k_B T_{\rm e,i, max}/ m_e}=0.005 c \ll v_0=0.2 c$.
The reason we refer to the maximal temperature
is because it actually varies across x-axis.
This is because as in \citet{Tsiklauri_2023} we keep
pressure balance by making $p_{\rm e,i}=n_{\rm e,i}(x)k T_{\rm e,i}(x)=
const$, i.e. $T_{\rm e,i}(x)=T_{\rm e,i, max}[n_0/n_{\rm e,i}(x)]
\propto 1/n_{\rm e,i}(x)$.
 With the magnetic field being uniform
$B_{\rm y0}=\bar B_0 B_{\rm MHD}= \bar B_0 \sqrt{\mu_0 m_e n_e} v_0$, given by
equation \ref{eq13}, this means that total initial 
pressure balance $p_{\rm e,i}+B^2/(2 \mu_0)=const$ is satisfied.
The factor $\bar B_0=0,0.5,1.0,5.0,7.5,10.0,15.0$
provides a variation the external magnetic field,
$B_{\rm y0}/B_{\rm MHD}=0,\;<1,\;1,\;\gg1$.
The electron and ion number densities are set as:\\
For the down-flows
\begin{equation}
n_{\rm e,i}(x)=
\begin{cases}
   n_0,& \text{if } (x < 0.25L_x) \text{ or } (x > 0.75L_x)\\
    10^{-2}n_0,              & \text{otherwise}
\end{cases}
\label{eq28}
\end{equation}
For the up-flow, vice versa
\begin{equation}
n_{\rm e,i}(x)=
\begin{cases}
   10^{-2}n_0,& \text{if } (x < 0.25L_x) \text{ or } (x > 0.75L_x)\\
    n_0,              & \text{otherwise}
\end{cases}
\label{eq29}
\end{equation}
The factor $10^{-2}$, while drops density to nearly zero, 
stops EPOCH from slowing down for numerical reasons. 
In EPOCH code physical quantities are in
SI units, so we fix $n_0=10^{15}$ particles per m$^{-3}$
typical of many collisionless astrophysical plasmas. 
The terms up-flow and down-flows only refer to
motions {\it figurally} up and down the $Oy$ axis, as
there is no gravity present in our simulations or calculations.
The plasma consists of electrons and protons
with the realistic mass ration $m_{\rm p}/m_{\rm e}=1836$.
Both electrons and ions are mobile throughout the simulation.
At $t=0$ electrons and ions velocities are set as described above i.e. 
a "sandwich" with three layers of plasma. This ensures
that initially there is {\it zero net current}.
We force the initial zero-net-current condition because when 
ions are stationary (these results are not included here),
strong oscillations with electron plasma frequency $\omega_{\rm pe}$ appear, e.g. in $(E_B-E_B(0))/E_p(0)$ and 
$(E_p-E_p(0))/E_p(0)$, Figures \ref{fig3} and \ref{fig5}, respectively.
Thus, zero-net-current condition is necessary to study a clear effect of
the external magnetic field on ESKHI, not marred by the said oscillations.
In EPOCH implementation we effectively load 4 plasma
species,  electron and  proton up-flows and down-flows as
specified by Eqs. \ref{eq28} and \ref{eq29}.
We use 200 particles per cell for each species. so 
in total, we have 
$4\times200\times200\times100=1.6\times10^7$
particles. One numerical run takes circa 40 minutes
on 12 processor cores.
In this paper we only show
snapshots of electron number density $n_e$ and
the generated by ESKHI magnetic field $B_z$ component 
normalized on $B_{\rm MHD}=({m_e  \omega_{\rm pe}}/{e}) 
({v_0}/{c})$. The length is normalized on $c/\omega_{\rm pe}$.
The end simulation time is set $150/\omega_{\rm pe}$. 

\begin{figure*}
\captionsetup{justification=raggedright,
singlelinecheck=false}
\begin{center}
  \makebox[\textwidth]{\includegraphics[width=0.99\textwidth]{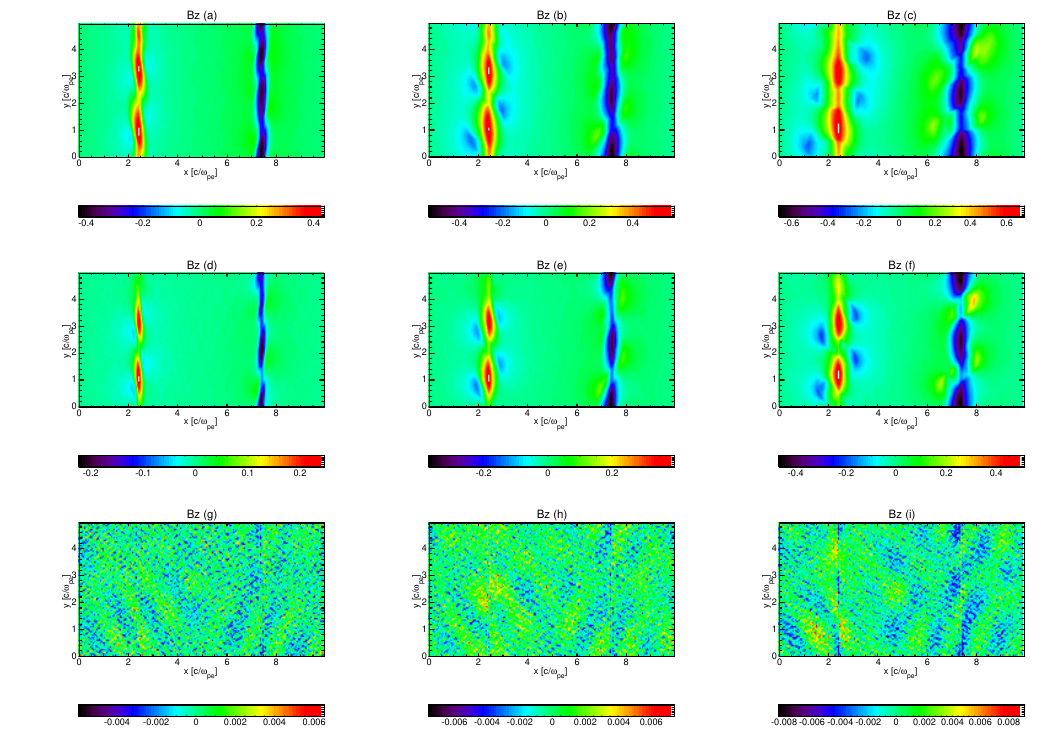}}
\end{center}
\caption{Snapshots of out-of-plane magnetic field $B_z$
generated by the ESKHI.
The numerical runs and the snapshot
times are in direct correspondence to Fig.\ref{fig1}.}
    \label{fig2}
\end{figure*} 

\begin{table}
\captionsetup{justification=raggedright,
singlelinecheck=false}
\caption{Table of numerical runs considered.
See text discussing Fig.\ref{fig3} for
the explanation of the notation used.}
\centering
\begin{tabular}{lccr} 
\hline
Run &  $\bar B_0=B_{\rm y0}/B_{\rm MHD}$ & $(\Gamma \omega_{\rm pe})_{\rm Diam}$=\texttt{r\_i[1]}& $(\Gamma \omega_{\rm pe})_{\rm Fit}$\\
\hline
0 & 0.0 & 0.3667 & 0.3663 \\
1 & 0.5 & 0.3138 & 0.3061 \\ 
2 & 1.0 & 0.2504 & 0.2629 \\ 
3 & 5.0 & 0.1340 & 0.1234 \\
4 & 7.5 & 0.0834 & 0.0927 \\
5 & 10.0 & 0.0690 & 0.0742 \\
6 & 15.0 & 0.0668 & 0.0531 \\
\hline
\end{tabular}
\label{t1}
\end{table}

\section{The Results}

In Fig.\ref{fig1} we show snapshots of electron
number density 
$n_{\rm e}$, which is the sum of down and up flowing electrons.
The row of panels (a), (b) and (c) show $n_{\rm e}/n_0$  for
Run 0 at times $t=20, \; 25, \; 30/\omega_{\rm pe}$.
The row of panels (d), (e) and (f) show $n_{\rm e}/n_0$ for
Run 2 at times $t=20, \; 25, \; 30/\omega_{\rm pe}$.
The row of panels (g), (h) and (i) show $n_{\rm e}/n_0$ for
Run 6 at twice the times $t=40, \; 50, \; 60/\omega_{\rm pe}$.
We gather from Fig.\ref{fig1} that in the case
zero external magnetic field (Run 0), the elongated, rotating
vortices are progressively generated.
We note significant similarities of panels corresponding
the Run 0 to a similar numerical run with the same number density
across the shear interface
from \citet{Alves_2014}, see their Figure 6.
The vortices start as narrow elongated flow structures with
under-dense cores $n_{\rm e}/n_0\simeq0.5$ and strongly over-dense
edges $n_{\rm e}/n_0\simeq2$. Note the values on the color bar.
As the time progresses from 20 to $30/\omega_{\rm pe}$
the vortex core-edge contrasts deepen even further and the width of vortices
grows. Such large values of under- and over-density indicates
strongly non-linear evolution of these ESKHI-generated vortices.
As the external magnetic field is increased (Run 2)
the values of under- and over-density in the vortices drop initially.
In the case of $\bar B_0=B_{\rm y0}/B_{\rm MHD}\gg1$ (Run 6)
vortices disappear altogether and only linear
amplitude waves can be seen generated in the vicinity
of the shear interfaces $x/L=0.25$ and 0.75.
As can be seen in Figure \ref{fig1}, panels (a)--(f),  
the simulation results
 are not affected
by the periodic boundary conditions used, because across x-axis
vortices never reach boundaries, while they simply leave and re-enter
at the top and bottom boundaries across y-axis.

We also note that in Figure \ref{fig1}, 
the electron number density has spatial 
variations in both the $x$- and $y$- directions. 
This is because the formed by ESKHI  vortices are rotating
in  $xOy$ plane. The theoretical model presented in the Appendix A
only considers $x$-variation for the perturbations. 
Strictly speaking, our theoretical model should also have
$y$-variation. However we retain only $x$-variation 
for {\it simplicity}, because, 
as explained in the conclusions, we only use the 
Eqs. \ref{eq21}-\ref{eq23}, and what follows from them 
just to {\it set the scale} of 
two-fluid (non-relativistic) MHD stability 
threshold magnetic field, given by Eqs. \ref{eq13} and \ref{eq14},
which we use in our PIC simulation as a relevant scale,
$B_{\rm MHD}$.

Fig.\ref{fig2} has a purpose to quantify 
how the out-of-plane magnetic field $B_z$ is
generated by the ESKHI.
We see in Fig.\ref{fig2} that in the case
zero external magnetic field (Run 0) 
DC (although with a corrugated shape) magnetic field $B_z$
is progressively generated
in the vicinity
of the shear interfaces $x/L=0.25$ and 0.75.
The values, judging from the color-bar, 
range $\pm (0.4-0.6) B_{\rm MHD}$ and they grow as the time progresses
from left to right panels.
Again, we mention that significant similarities can be seen
in  panels corresponding
the Run 0 to a similar numerical run with the same number density
across the shear interface
from \citet{Alves_2014}, see their Figure 7.
For the further increased (increased from zero) 
external magnetic field, for Run 2,
 we see that the generated values of $B_z$
are about factor of two smaller compared to Run 0.
Also gaps in the generated DC field appear,
as these structures further narrow down across the
shear interfaces at $x/L=0.25$ and 0.75.
For Run 6, the values of $B_z$ drop to 0.005 near
the shear interfaces at $x/L=0.25$ and 0.75.

\begin{figure}
\captionsetup{justification=raggedright,
singlelinecheck=false}
\includegraphics[width=\columnwidth]{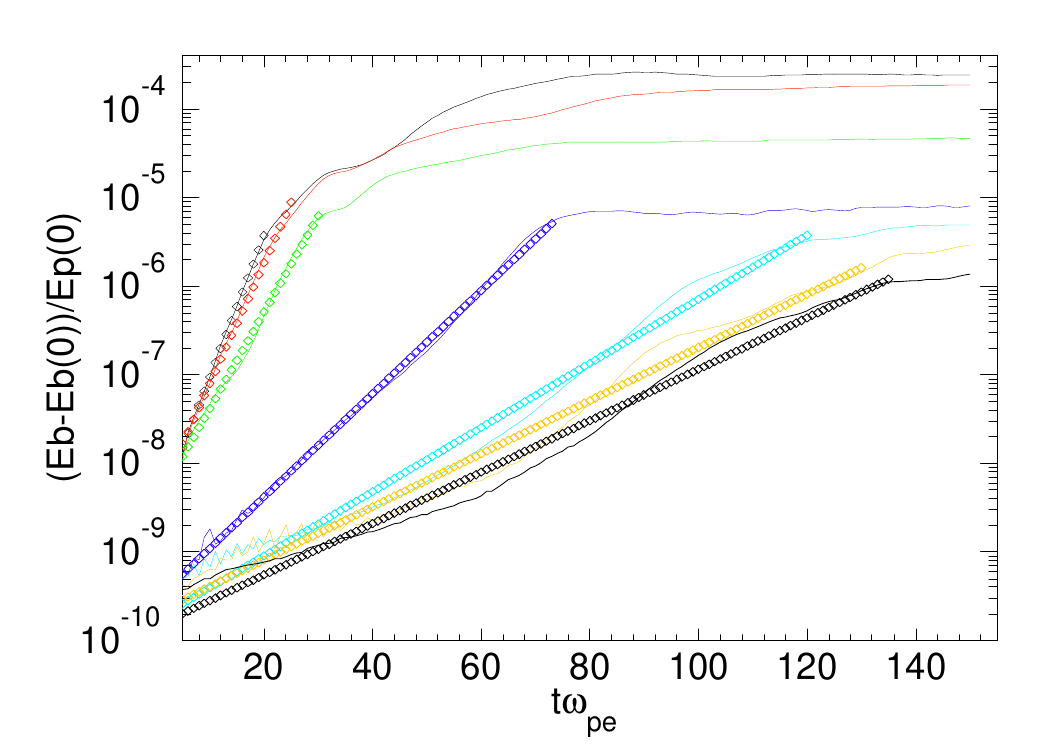}
    \caption{Black, red, green, blue, cyan, gold and thick solid curves
show $(E_B-E_B(0))/E_p(0)$ using numerical
simulation data from Run0 -- Run 6 using  Eq.\ref{eq30}.
The open diamonds with the same multi-colors are
showing the fit using Eq.\ref{eq31}. }
    \label{fig3}
\end{figure}
In Fig.\ref{fig3} we show 
the time evolution of {\it perturbation}
equipartition energy $(E_B-E_B(0))/E_p(0)$.
It is crucial that we subtract $E_B(0)$
because we want to separate the pre-existing
magnetic field energy contribution from
the ESKHI-generated magnetic field energy.
We calculate this quantity in the following
way: at every time step we load the data
and calculate the following quantity:
\begin{equation}
\begin{split}
&\frac{E_B(t)-E_B(0)}{E_p(0)}=\\
&\iint \frac{b_x(x,y,t)^2+[b_y(x,y,t)-B_{y0}]^2+b_z(x,y,t)^2}{2 \mu_0E_p(0)}dxdy=\\
&\frac{1}{2\mu_0E_p(0)}
\sum_{i=1}^n
[b_x(x_i,y_i,t)^2+[b_y(x_i,y_i,t)-B_{y0}]^2+\\
&b_z(x_i,y_i,t)^2]\Delta x\Delta y.
\label{eq30}
\end{split}
\end{equation}
Note that in Eq.\ref{eq30} the numerical integration
is done by the midpoint rule (also known as the rectangle rule).
In Fig.\ref{fig3} black, red, green, blue, cyan, gold and thick solid curves
show $(E_B-E_B(0))/E_p(0)$ using numerical
simulation data from Run0 -- Run 6 using  Eq.\ref{eq30}.
We gather that ESKHI rapidly grows the magnetic
perturbation equipartition energy on the time scales of 30 to 
$120 /\omega_{\rm pe}$ depending on the strength
of the background external magnetic field.
This exponential growth phase is followed by a plateau
which as explained by \citet{Alves_2014} is due
to the {\it generated} by ESKHI magnetic field
component $B_z$ blocks the flow of electrons across the
shear interfaces at $x/L=0.25$ and 0.75.
We also see that as $\bar B_0=B_{\rm y0}/B_{\rm MHD}$
increases from 0 to 15 the growth rate of ESKHI decreases
considerably.
The open diamonds of the same color, as stated above,
show our fit using the following method.
We use Interactive Data Language (IDL)'s built-in function
called 
\texttt{poly\_fit} to calculate $r[0]_i$ and $r[1]_i$
that appear in Equation \ref{eq31}, where
$i=0,\ldots,6$ for each of the Run0 -- Run 6.
Effectively, this function fits a 1st order polynomial
to natural logarithm of $(E_B-E_B(0))/E_p(0)$
with $r[0]_i$ and $r[1]_i$ being the 0th and and 1st 
order fit coefficients to the polynomial.
This fit then enables to
plot with open diamonds showing data fit using Eq. \ref{eq31}: 
\begin{equation}
e^{r_i[0]}\times e^{r_i[1] \omega_{\rm pe}  t}.
\label{eq31}
\end{equation}
The values of $r_i[1]$ are quoted as the 3rd
column of Table \ref{t1}. Effectively these
values are the growth rates of ESKHI 
$(\Gamma \omega_{\rm pe})_{\rm Diam}$=\texttt{r\_i[1]}
shown with 
multi-color open diamonds in 
Fig.\ref{fig3}. The "Diam" is abbreviation for diamonds from Fig.\ref{fig3}.

\begin{figure*}
\captionsetup{justification=raggedright,
singlelinecheck=false}
    \centering
    \begin{subfigure}[t]{0.45\textwidth}
        \centering
        \includegraphics[width=\linewidth]{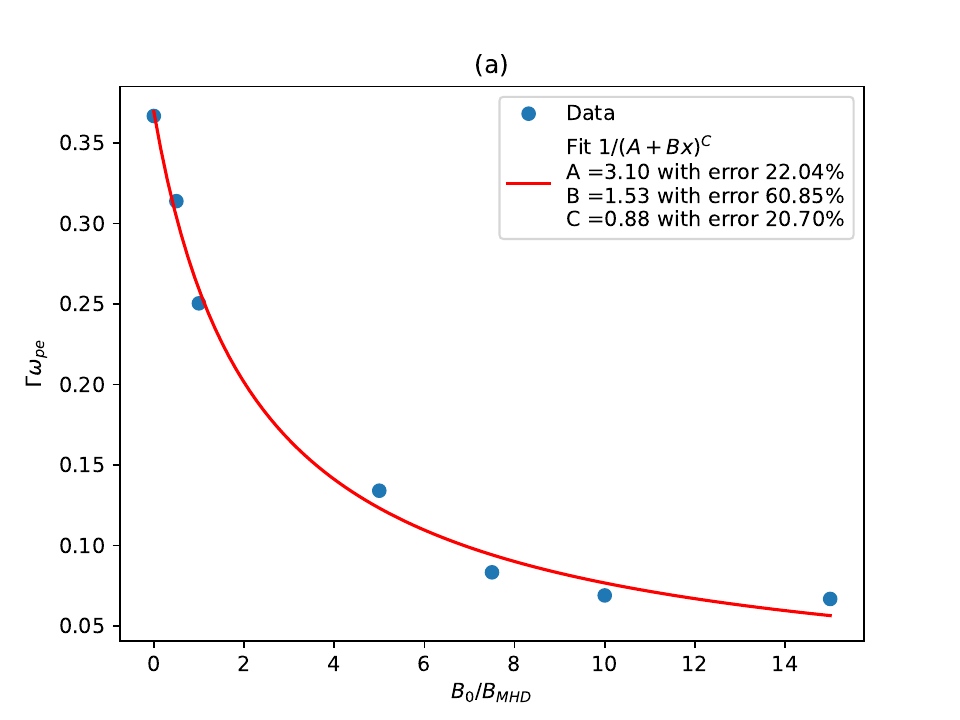} 
    \end{subfigure}
    \hfill
    \begin{subfigure}[t]{0.45\textwidth}
        \centering
        \includegraphics[width=\linewidth]{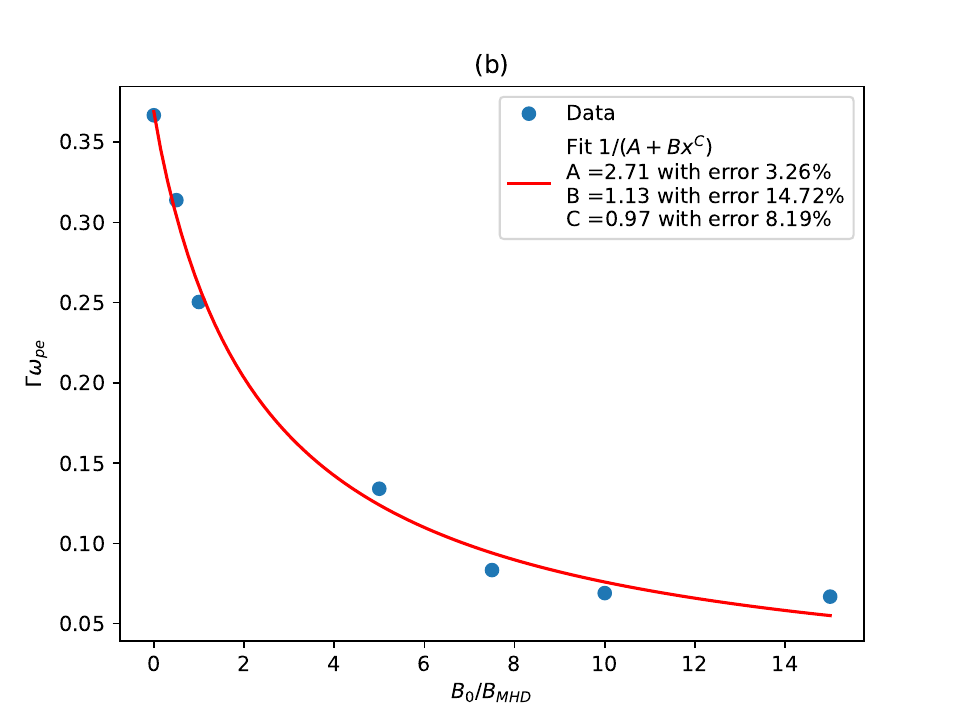} 
    \end{subfigure}
    \begin{subfigure}[t]{0.45\textwidth}
        \centering
        \includegraphics[width=\linewidth]{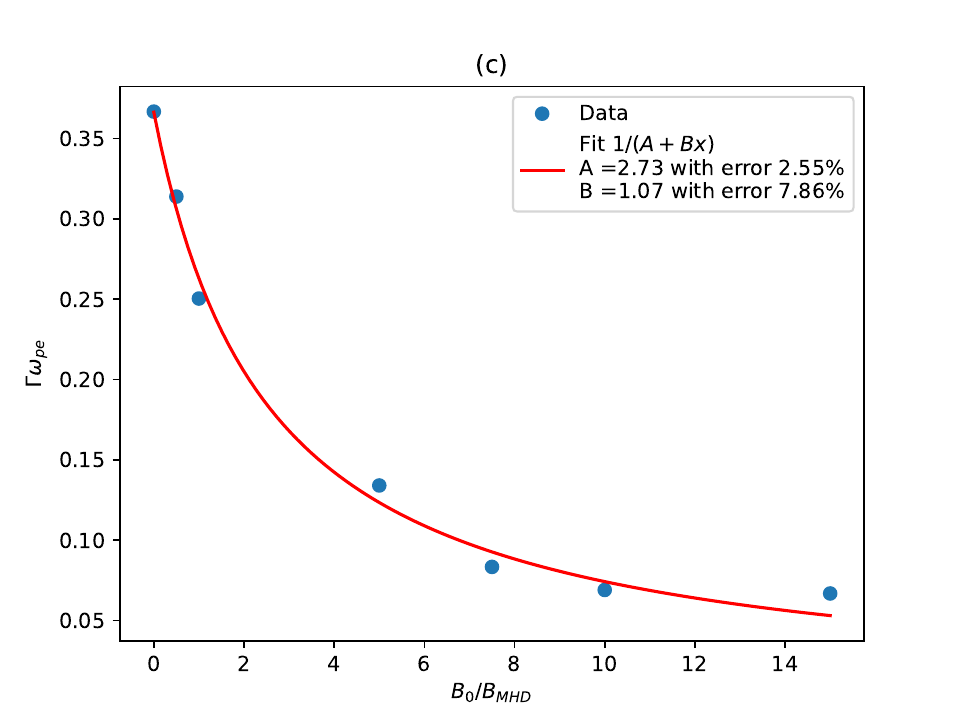} 
    \end{subfigure}
    \hfill
    \begin{subfigure}[t]{0.45\textwidth}
        \centering
        \includegraphics[width=\linewidth]{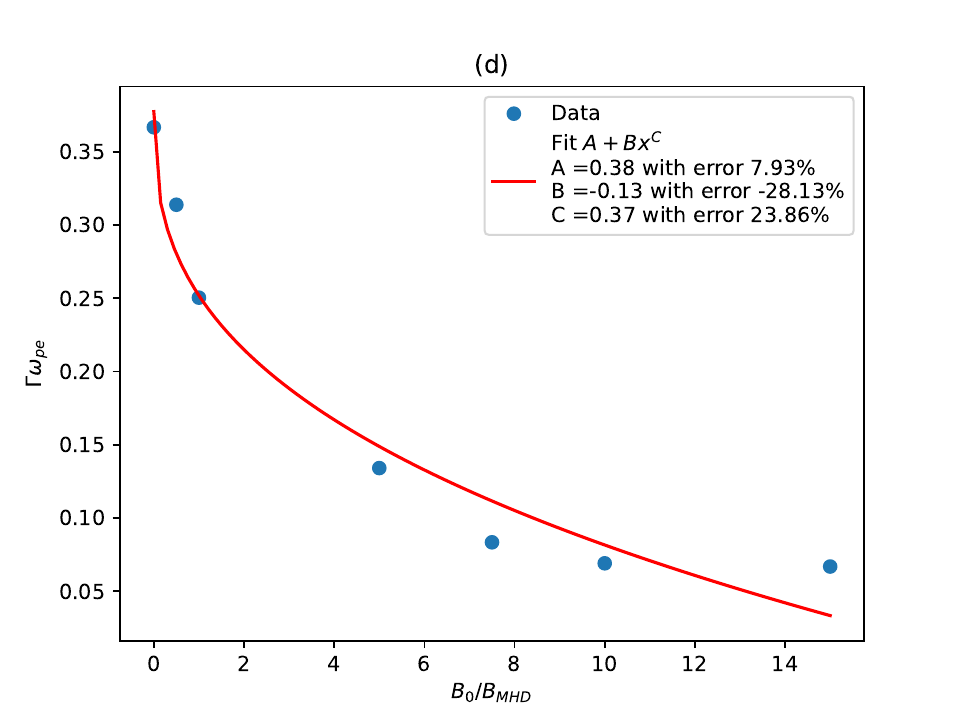} 
    \end{subfigure}
    
    \begin{subfigure}[t]{0.45\textwidth}
        \centering
        \includegraphics[width=\linewidth]{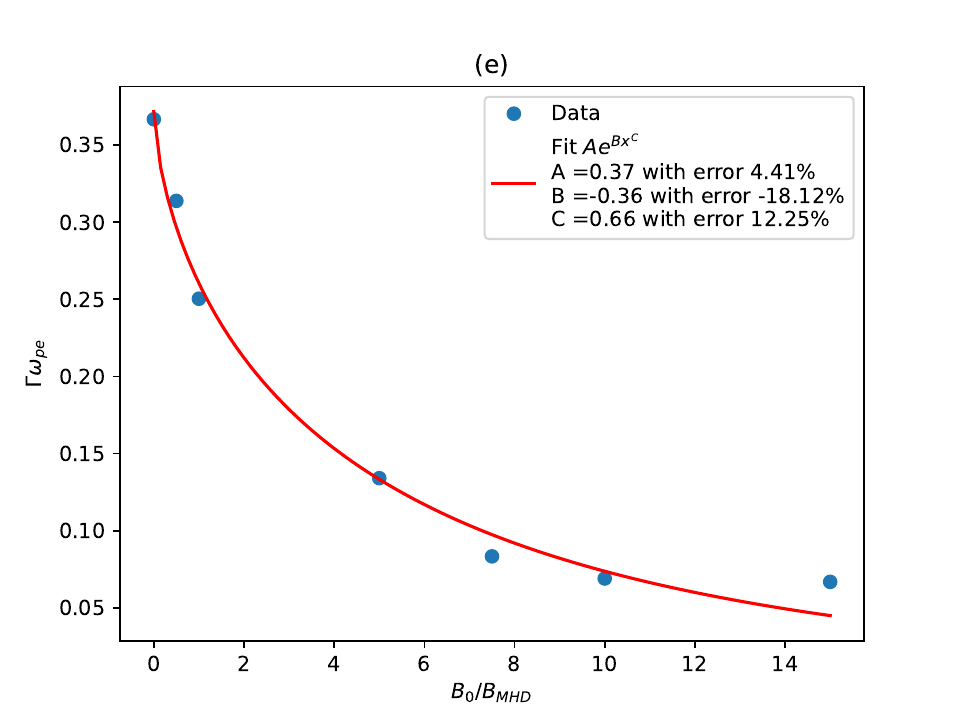} 
    \end{subfigure}
    \hfill
    \begin{subfigure}[t]{0.45\textwidth}
        \centering
        \includegraphics[width=\linewidth]{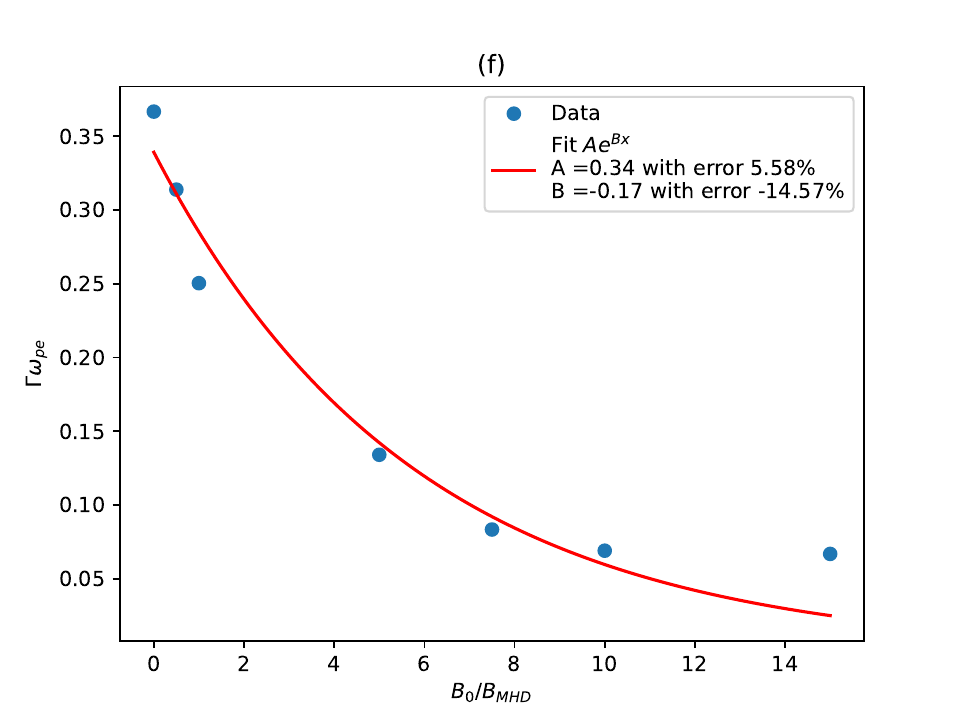} 
    \end{subfigure}    
 \caption{Various plausible functions
 fitted to $(\Gamma \omega_{\rm pe})_{\rm Diam}$ values i.e.
the best possible functional fit to the
3rd column of Table \ref{t1}.}
    \label{fig4}
\end{figure*}

In Fig.\ref{fig4}  we would like to deduce 
the effect of external magnetic field on ESKHI in a {\it functional 
dependence form}.
In other words, we would like to know what function can be fitted
to $(\Gamma \omega_{\rm pe})_{\rm Diam}$.
Various plausible functions were attempted.
Only a small fraction of fit functions is shown
in panels (a)--(f) in Fig.\ref{fig4}.
We gather from Fig.\ref{fig4} that in panel (c)
we have the best fit with the smallest errors.
Thus we conclude that 
the best fit is hyperbolic, 
$\Gamma(B_0)\omega_{\rm pe}=\Gamma_0\omega_{\rm pe}/(A+B\bar B_0)$, where $\Gamma_0\omega_{\rm pe}=1/\sqrt{8}=0.35$ is the 
electron scale KHI growth rate without external magnetic field
and $\bar B_0=B_0/B_{\rm MHD}$ is the ratio of
external and  two-fluid MHD stability threshold magnetic field.
The hyperbolic fit numerical values of the growth rate
are quoted for reference as the 4th column in Table \ref{t1}.
Indeed, as can be deduced both from Fig. \ref{fig4}(c) and the 3d and 
4th columns in Table \ref{t1}, the fit graphically and numerically is 
rather good.
The factor, which is the x-axis in  Fig.\ref{fig4}
$\bar B_0=0,0.5,1.0,5.0,7.5,10.0,15.0$
provides a variation the external magnetic field.
Paradoxically, for the 
growth rate predicted by Eq.(26) which is supposed to
be the best fit, the fit even does not converge due to large errors.

\begin{figure}
\captionsetup{justification=raggedright,
singlelinecheck=false}
\begin{center}
 \includegraphics[width=\columnwidth]{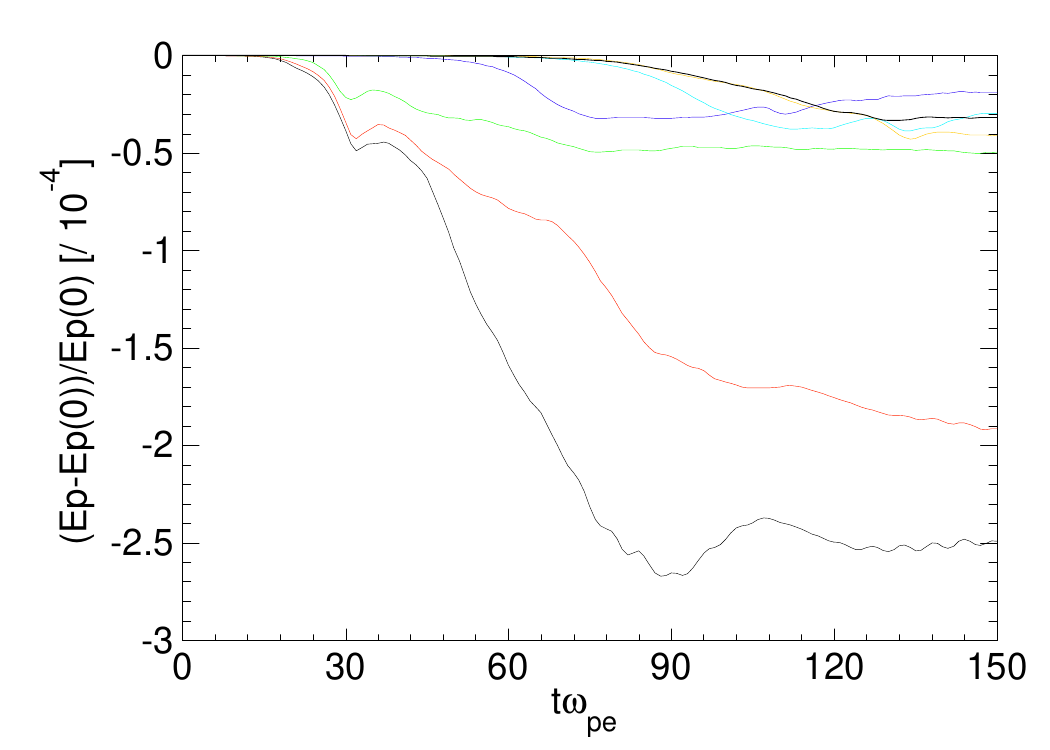}
\end{center}
\caption{Time evolution of 
particle perturbation kinetic energy $(E_p-E_p(0))/E_p(0)$
using numerical
simulation data from Run 0 -- Run 6 with the same multi-color
curves as in Fig.\ref{fig3}}
\label{fig5}
\end{figure}
In Fig.\ref{fig5} we show 
the time evolution of 
particle {\it perturbation} kinetic energy $(E_p-E_p(0))/E_p(0)$
using numerical
simulation data from Run0 -- Run 6 with the same multi-color
curves as in Fig.\ref{fig3}.
We deduce two observations from Fig.\ref{fig5}:
(i) The magnetic field generated by ESKHI comes at an expense
of reduction in particle kinetic energy;
(ii) The strongest reduction in the kinetic energy is seen
for the case of Run 0, a zero background external magnetic field when
the ESKHI growth rate is the largest.
As the external magnetic field is increased, this arrests the flow
of electrons, across the shear
interfaces and we see lesser and lesser reduction in the kinetic
energy, as a result of this.

\begin{figure}
\captionsetup{justification=raggedright,
singlelinecheck=false}
\begin{center}
  \includegraphics[width=\columnwidth]{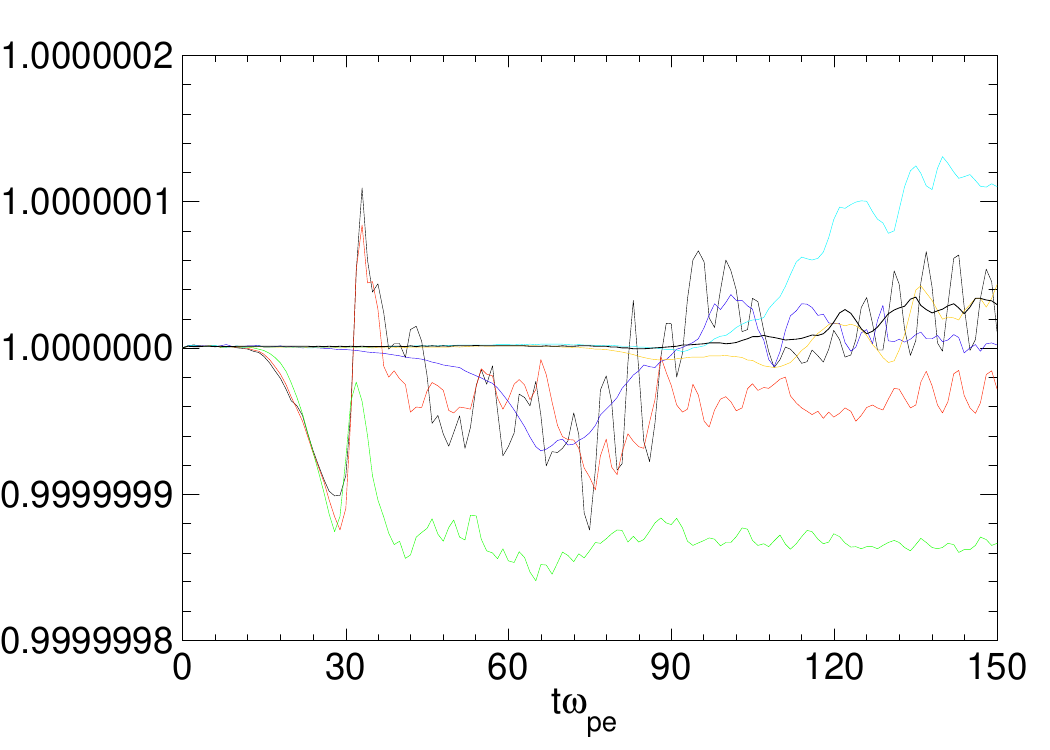}
\end{center}
\caption{Similar to Figure \ref{fig5} but now for 
    the total energy.}
\label{fig6}
\end{figure}
In Fig.\ref{fig6} we plot the total energy 
$(E_p+E_f)/(E_p(0)+E_f(0))$ in a
similar manner to Figure \ref{fig5}.
Note that $E_f$ denotes the total EM field energy,
automatically calculated by EPOCH as 
\texttt{data.TOTAL\_FIELD\_ENERGY}, which includes
contribution from both magnetic and electric fields,
while in Fig.\ref{fig3} we had to calculate perturbation
equipartition energy $(E_B-E_B(0))/E_p(0)$
manually, using Eq.\ref{eq30}.
We gather from Fig.\ref{fig6} that the
total energy conservation in all our 
EPOCH numerical runs is superb and the relative 
errors are contained
within a small margin of
$\pm 0.0000001$ i.e. $\pm 0.00001\%$.

\section{Conclusions}

ESKHI can be of importance in many collisionless plasmas e.g. 
when solar wind interacts with planetary magnetospheres, 
magnetic field is generated in AGN jets, or shocks and flows
found in Gamma Ray Bursts \citep{gruzinov2008,Alves_2012,Grismayer_2013,Alves_2014,Alves_2015}. 
The aim of this paper is
to study the effect of external, background magnetic field 
on ESKHI. Thus we use particle-in-cell, fully electromagnetic 
plasma simulation as the main tool for this purpose. 
This study finds that in the kinetic regime, 
the presence of external magnetic field reduces 
the growth rate of the instability -- a result similar to
the well-known analogue --  magnetohydrodynamic KHI.
While, in MHD there is  known threshold 
magnetic field for KHI stabilization as first shown
by \citet{Michael_1955}, for ESKHI
this is yet to be determined by an appropriate analytical
calculation that would extend approach used in \citep{Alves_2014,Alves_2015} 
by adding the external, background magnetic field.
Such calculation is rather complex if all three components
of velocity $v_x,v_y,v_z$ are considered.
Note that in EPOCH in all 1.5, 2.5, 3D versions all three
components of velocity are always present.
Instead, for the purposes of this paper, 
we only use the calculations based on
Eqs. \ref{eq21}-\ref{eq23}, and their outcomes 
to set the scale of 
two-fluid MHD stability threshold magnetic field, derived
in Appendix A and given by Eq. \ref{eq14}.
As it stands, without a fully kinetic analytical expression 
for the growth rate,
we decided to use several numerical simulation runs to find
an empirical
dependence of ESKHI growth rate, $\Gamma(B_0)\omega_{\rm pe}$, 
on the strength of applied external magnetic field. 
Our results show that the best fit is hyperbolic, 
$\Gamma(B_0)\omega_{\rm pe}=\Gamma_0\omega_{\rm pe}/(A+B\bar B_0)$.
We note an urgent need for an analytical theory 
to back up the said growth rate dependence on
the external magnetic field.
The first peculiar and important result that follows from
our study is that in astrophysical objects where a
strong magnetic field pre-exists, the generation
of an additional magnetic field by the ESKHI is
suppressed. The latter suggests that 
with this, the Nature provides a 
"safety valve" -- natural protection 
not to "over-generate" magnetic field.
The second peculiar result is that
we show
that two-fluid (nonrelativistic) MHD threshold magnetic field,
calculated in Appendix A, 
(Eq.\ref{eq13} or equally Eq.\ref{eq14})
is the same (up to a factor of $\sqrt{\gamma_0}\approx~1$)
as the DC saturation magnetic field, predicted 
by the fully kinetic theory (Eq.\ref{eq9})
established by \citet{Grismayer_2013,Alves_2014}. 

This work was complete when the 
author has become aware (G.P. Zank, private communication)
of \citet{Che_2023}. The calculation shown in our Appendix A
and also our Eq.\ref{eq27}
appear to be similar to that of \citet{Che_2023}.
However, it is clear that neither 
Eq.(41) of \citet{Che_2023}
nor our Eqs.\ref{eq24}--\ref{eq26}
agree with the 
best fit, 
$\Gamma(B_0)\omega_{\rm pe}=\Gamma_0\omega_{\rm pe}/(A+B\bar B_0)$,
established by the present work.
This disagreement means that
a new analytical calculation for the same number density across the shear interface as in \citet{Alves_2014} with {\it added} 
external magnetic field $B_0$ along the shear interface is needed.
Further, the results from 
Table \ref{t1} can be used to check  the validity
of the said growth rate yet to be analytically determined.  
Such analytical theory to back up the established here 
by PIC simulation growth rate dependence on
external magnetic field is urgently needed.
We stress again that, we only use the calculations based on
Eqs. \ref{eq21}-\ref{eq23}, and their outcomes 
just to set the scale of 
two-fluid (non-relativistic) MHD stability 
threshold magnetic field, given by Eqs. \ref{eq13} and \ref{eq14},
used in our PIC simulation as a relevant scale,
$B_{\rm MHD}$.

\section*{Acknowledgements}
Author would like to thank an anonymous 
referee whose comments improved this manuscript.

\section*{Data Availability}
The data that support the findings of this study are available
from the corresponding author upon reasonable request.

\appendix
\section{Appendixes}

Starting with the equations of two-fluid MHD with stationary
ions \ref{eq21}--\ref{eq23}, we consider
two magnetized electron flows with properties
$\rho_{e1}$, $p_{e1}$, $U_{e1}$, $B_1$ for $y<0$ and
$\rho_{e2}$, $p_{e2}$, $U_{e2}$, $B_2$ for $y>0$
with the interface at $y=0$. We only consider $xOy$ plane.
This calculation extends that of \citet{Michael_1955}
two-fold: (i) we consider situation where bulk
flow velocity is replaced by electron velocity, $\vec v\to\vec v_e$ and
$\rho \to \rho_{e}$; 
and (ii) the densities across the interface  are different.
For $y<0$ we have the following linearized
 equations in the component form:
\begin{equation}
\frac{\partial b_x}{\partial t}+U_{e1}\frac{\partial b_x}{\partial x}
=B_1 \frac{\partial v_{ex}}{\partial x}, 
\label{eqa1}
\end{equation}
\begin{equation}
\frac{\partial b_y}{\partial t}+U_{e1}\frac{\partial b_y}{\partial x}
=B_1 \frac{\partial v_{ey}}{\partial x}, 
\label{eqa2}
\end{equation}
\begin{equation}
\frac{\partial v_{ex}}{\partial t}+U_{e1}\frac{\partial v_{ex}}{\partial x}
=-\frac{1}{\rho_{0e1}} \frac{\partial p_{e1}}{\partial x}, 
\label{eqa3}
\end{equation}
\begin{equation}
\frac{\partial v_{ey}}{\partial t}+U_{e1}\frac{\partial v_{ey}}{\partial x}
=-\frac{1}{\rho_{0e1}} \frac{\partial p_{e1}}{\partial y}+
\frac{B_1}{\mu_0\rho_{0e1}}\left(\frac{\partial b_y}{\partial x}-
\frac{\partial b_x}{\partial y}\right), 
\label{eqa4}
\end{equation}
\begin{equation}
\frac{\partial b_x}{\partial x}+
\frac{\partial b_y}{\partial y}=0, \;\;\;\;\;\;
\frac{\partial v_{ex}}{\partial x}+
\frac{\partial v_{ey}}{\partial y}=0,
\label{eqa5}
\end{equation}
where $v_{ex,ey}$ and $b_{x,y}$
are the perturbations of background values of $U_{e1}$ and $B_1$.
Next, we substitute in Eqs. \ref{eqa1}--\ref{eqa5} 
Fourier ansatz of the form $f=\tilde f e^{i(kx+\omega t)}$
and omit the tilde signs:
\begin{equation}
i(\omega+k U_{e1})b_x=B_1 ik v_{ex},
\label{eqa6}
\end{equation}
\begin{equation}
i(\omega+k U_{e1})b_y=B_1 ik v_{ey},
\label{eqa7}
\end{equation}
\begin{equation}
i(\omega+k U_{e1})v_{ex}=-\frac{ik}{\rho_{0e1}} p_{e1},
\label{eqa8}
\end{equation}
\begin{equation}
i(\omega+k U_{e1})v_{ey}=-\frac{1}{\rho_{0e1}} 
\frac{\partial p_{e1}}{\partial y}+
\frac{B_1}{\mu_0\rho_{0e1}}\left(ik b_y-
\frac{\partial b_x}{\partial y}\right),
\label{eqa9}
\end{equation}
\begin{equation}
ik b_x+\frac{\partial b_y}{\partial y}=0,
\label{eqa10}
\end{equation}
\begin{equation}
ik v_{ex}+\frac{\partial v_{ey}}{\partial y}=0.
\label{eqa11}
\end{equation}
Eqs. \ref{eqa6}--\ref{eqa11} can be combined
into one master equation for $v_{ey}$:
\begin{equation}
\left(\frac{B_1^2 k^2}{\mu_0\rho_{0e1}}-
(\omega+kU_{e1})^2\right)
\left(k^2 v_{ey}-\frac{\partial^2 v_{ey}}{\partial y^2}\right)=0.
\label{eqa12}
\end{equation}
Following similar approach to
\citet{Michael_1955}, from Eq. \ref{eqa12}
we have a solution
for $y<0$ (medium 1)  $v_{ey1}=C_{11}e^{ky}+C_{12}e^{-ky}$.
For a solution which approaches zero as $y \to -\infty$ we have
$v_{ey1}=C_{11}e^{ky}$.
Likewise, in the region
with $y>0$ (medium 2), 
for a solution which approaches zero as $y \to \infty$ we have  
$v_{ey2}=C_{22}e^{-ky}$. 

Let $\xi(x,t)$ be displacement of the interface satisfying
a condition $\xi \ll 2 \pi / k=\lambda$.
As in \citet{Michael_1955},
based on the definition
$v_{ey1}=\partial \xi / \partial t+U_{e1}{\partial \xi}/{\partial x}$
and
$v_{ey2}=\partial \xi / \partial t+U_{e2}{\partial \xi}/{\partial x}$,
 we demand continuity 
of ${\partial \xi}/{\partial t}$:
\begin{equation}
\frac{\partial \xi}{\partial t}=
v_{ey1}-U_{e1}\frac{\partial \xi}{\partial x}=
v_{ey2}-U_{e2}\frac{\partial \xi}{\partial x},
\label{eqa13}
\end{equation}
which at the interface $y=0$ yields
\begin{equation}
(\omega +k U_{e1}) C_{11}=(\omega +k U_{e2}) C_{22}.
\label{eqa14}
\end{equation}
We also have to fulfill the condition for the pressure
balance
$p_e+B^2/(2\mu_0)=const$, substituting $p_e=p_{e0}+p_{e1}$
and $B=B_{1,2}+b_{1,2}$,
 at the interface, at a linear order. Omitting
the quadratic terms, at the interface $y=0$  we obtain:
\begin{equation}
p_{e1}+\frac{B_1 b_{x1}}{\mu_0}=
p_{e2}+\frac{B_2 b_{x2}}{\mu_0}.
\label{eqa15}
\end{equation}
Substituting the following quantities for medium 1
\begin{equation}
p_{e1}=\frac{-i(\omega+k U_{e1})\rho_{e1}C_{11}}{k},
\label{eqa16}
\end{equation}
\begin{equation}
b_{x1}=\frac{ik B_1C_{11}}{(\omega+k U_{e1})},
\label{eqa17}
\end{equation}
and similar expressions for medium 2
\begin{equation}
p_{e2}=\frac{i(\omega+k U_{e2})\rho_{e2}C_{22}}{k},
\label{eqa18}
\end{equation}
\begin{equation}
b_{x2}=-\frac{ik B_2C_{22}}{(\omega+k U_{e2})},
\label{eqa19}
\end{equation}
into Eq. \ref{eqa15} and
 after multiplying both sides by $k$, we obtain:
\begin{equation}
\begin{split}
C_{11}&\left(-(\omega+k U_{e1})\rho_{e1}+
\frac{B_1^2k^2}{\mu_0(\omega+k U_{e1})}\right)=\\
C_{22}&\left((\omega+k U_{e2})\rho_{e2}-
\frac{B_2^2k^2}{\mu_0(\omega+k U_{e2})}\right).
\label{eqa20}
\end{split}
\end{equation}
The next step is to make use of Eq. \ref{eqa14}
to obtain:
\begin{equation}
\begin{split}
&\frac{\left(-(\omega+k U_{e1})\rho_{e1}+
\frac{B_1^2k^2}{\mu_0(\omega+k U_{e1})}\right)}{(\omega+k U_{e2})}=\\
&\frac{\left((\omega+k U_{e2})\rho_{e2}-
\frac{B_2^2k^2}{\mu_0(\omega+k U_{e2})}\right)}{(\omega+k U_{e1})}.
\label{eqa21}
\end{split}
\end{equation}
Further, simple algebra leads us to a quadratic
equation
\begin{equation}
\begin{split}
&(\rho_{e1}+\rho_{e2})\omega^2+2k(U_{e1}\rho_{e1}+
U_{e2}\rho_{e2})\omega+\\
&\left[k^2\left(U_{e1}^2\rho_{e1}+
U_{e2}^2\rho_{e2}\right)+\frac{B_1^2k^2}{\mu_0}+\frac{B_2^2k^2}{\mu_0}\right]=0.
\label{eqa22}
\end{split}
\end{equation}
Eq. \ref{eqa22} is a quadratic equation with respect to $\omega$,
which has the solution given by
Eq. \ref{eq24}.



\bibliography{ms2024-0224}
\bibliographystyle{aasjournal}
\end{document}